\documentclass[%
prb,%
twocolumn,
aps,%
showpacs,%
a4paper,%
superscriptaddress%
]{revtex4-1}

\usepackage{graphicx}
\usepackage{bm}
\usepackage{subfigure}
\usepackage{amsmath}
\usepackage{amssymb}
\usepackage{float}
\usepackage[latin1]{inputenc}
\usepackage[T1]{fontenc}
\usepackage[USenglish]{babel}
\usepackage{microtype}
\usepackage{nicefrac}
\usepackage{todonotes}
\newcommand{\figref}[1]{Fig.~\ref{#1}}
\newcommand{\intextcite}[1]{Ref.~\citenum{#1}}

\newcommand{\unit}[1]{~\mathrm{#1}}
\newcommand{\rangstrom}{~\mbox{\AA}^{-1}}

\renewcommand{\Im}{%
\mathrm{Im}
}

\newcommand{\bscco}{%
Bi$_2$Sr$_2$CaCu$_2$O$_{8+\delta}$%
}

\newcommand{\naccox}{%
Ca$_{2-x}$Na$_x$CuO$_2$Cl$_2$%
}
\newcommand{\naccoc}{%
Ca$_{1.95}$Na$_{0.05}$CuO$_2$Cl$_2$%
}
\newcommand{\naccod}{%
Ca$_{1.9}$Na$_{0.1}$CuO$_2$Cl$_2$%
}
\newcommand{\scoc}{%
Sr$_{2}$CuO$_2$Cl$_2$%
}
\newcommand{\cuopl}{%
CuO$_2$ plane%
}

\newcommand{\direction}[1]{%
\ensuremath{\left[#1\right]}%
}
\newcommand{\intextlf}{%
$\Im(-1/\epsilon(\omega,\bm{q}))$%
}
\newcommand{\titleofpaper}{%
Angular Dependent Spectral-Weight Transfer and Evidence for Symmetry Broken In-Plane Charge Response in \naccod
}
\usepackage{hyperref}
\hypersetup{pdftitle = {\titleofpaper}, pdfauthor = {Roman Schuster}, colorlinks,citecolor = blue, urlcolor = blue}
\usepackage[all]{hypcap}
\begin{document}
\title{\titleofpaper}
\author{R. Schuster}
\affiliation{IFW Dresden, Institute for Solid State Research, P.O. Box 270116, D-01171 Dresden, Germany} 
\author{S. Pyon}
\altaffiliation[Current address: ]{Department of Physics, Faculty of Science, Okayama University, Okayama 700-8530, Japan}
\affiliation{Department of Physics, University of Tokyo, Hongo, Tokyo 113-0033, Japan}
\author{M. Knupfer}
\affiliation{IFW Dresden, Institute for Solid State Research, P.O. Box 270116, D-01171 Dresden, Germany}
\author{M. Azuma}
\affiliation{Materials and Structures Lab., Tokyo Institute of Technology, Yokohama, Kanagawa 226-8503, Japan}
\author{M. Takano}
\affiliation{Institute for Integrated Cell-Material Sciences, Kyoto University, Sakyo-ku, Kyoto 606-8501, Japan} 
\author{H. Takagi}
\affiliation{Department of Physics, University of Tokyo, Hongo, Tokyo 113-0033, Japan}
\affiliation{Magnetic Materials Laboratory, RIKEN Advanced Science Institute, Wako, Saitama 351-0198, Japan}
\affiliation{Inorganic Complex Electron Systems Research Team, RIKEN Advanced Science Institute, Wako, Saitama 351-0198, Japan}
\author{B. Büchner}
\affiliation{IFW Dresden, Institute for Solid State Research, P.O. Box 270116, D-01171 Dresden, Germany}
\affiliation{Institute for Solid-State Physics, Department of Physics, TU Dresden, D-01062 Dresden, Germany}
\date{\today}

\begin{abstract}
We report about the energy and momentum dependent charge response in \naccod\ employing electron energy-loss spectroscopy. Along the diagonal of the Brillouin zone (BZ) we find a plasmon peak---indicating the presence of metallic states in this momentum region---which emerges as a consequence of substantial spectral-weight transfer from excitations across the charge-transfer (CT) gap and is the two-particle manifestation of the small Fermi pocket or arc observed  with photoemission in this part of the BZ. In contrast, the spectrum along the \direction{100}-direction is almost entirely dominated by CT excitations, reminiscent of the insulating parent compound. We argue that the observed polarization dependent shape of the spectrum is suggestive of a breaking of the underlying tetragonal lattice symmetry, possibly due to fluctuating nematic order in the charge channel. In addition we find the plasmon bandwidth to be suppressed compared to optimally doped cuprates.
\end{abstract}
\pacs{71.30.+h, 73.20.Mf, 74.72.Jt, 79.20.Uv}
\maketitle

\section{Introduction}
Cuprates are still the paradigm for many topics in condensed matter physics that cannot be reconciled within the conventional Fermi liquid approach. At the heart of the research on their electronic structure lies the question how the insulating parent compounds develop into high-temperature superconductors and finally into ordinary metals upon doping.

It is well established that, upon introducing charges into the \cuopl\ cuprates undergo a metal-insulator transition (MIT) which is seen in various transport and spectroscopic experiments \cite{Imada_Rev.Mod.Phys._1998_v70_p1039}. This MIT, however, appears to be rather peculiar because additional charges are believed to aggregate in particular patterns termed stripes or checkerboards \cite{Vojta_Adv.Phys._2009_v58_p699}, a phenomenon that is also observed in other transition-metal systems like the nickelates \cite{Tranquada_Phys.Rev.Lett._1994_v73_p1003}. In general, the charge order is accompanied by a corresponding structure in the spin sector \cite{Tranquada_Nature_1995_v375_p561}. In this respect the very recent reports about the existence of charge-only ordering  in $R$Ba$_2$Cu$_3$O$_{6+x}$ ($R=$Y,Nd) \cite{Ghiringhelli_Science_2012_v337_p821,Chang__2012_v_p} are remarkable. 

The \naccox\ system exhibits charge order in the form of bond-centered stripes or nematic behavior as revealed by scanning-tunneling microscopy (STM) experiments \cite{Hanaguri_Nature_2004_v430_p1001,Kohsaka_Science_2007_v315_p1380,Kohsaka_NatPhys_2012_v8_p534}. Naturally, the role of those inhomogeneities for the superconductivity or more generally for the electronic properties of underdoped cuprates is under strong debate. In particular it is known that in the La-based families there is a significant suppression of the superconducting transition temperature for $x=0.125$ where the stripe order is most robust \cite{Moodenbaugh_Phys.Rev.B_1988_v38_p4596,Tranquada_Nature_1995_v375_p561}. While the  spin dynamics in this region of the phase diagram is well established by inelastic neutron scattering (see e.g. \cite{Birgeneau_J.Phys.Soc.Jpn._2006_v75_p111003}) the corresponding behavior for excitations in the charge channel as a function of energy \emph{and} momentum is, though of urgent interest for a complete characterization of the collective-mode spectrum, only scarcely explored so far.

Inelastic electron scattering also known as electron energy-loss spectroscopy (EELS) in transmission is a well established and bulk-sensitive tool to investigate the energy and momentum dependence of the charge response in solids as its cross-section is directly proportional to  \intextlf---with $\epsilon(\omega,\bm{q})$ the complex dielectric function of the sample---and therefore allows to investigate the dynamics of collective charge excitations  \cite{Schnatterly1979}. 

Here we employ EELS to study the charge dynamics in the system \naccod. We find an angular dependent transfer of spectral weight between the excitations across the charge-transfer (CT) gap and those associated with the metallic state. The result of this is a ``nodal metal'' that is observed as the truncated Fermi surface (FS) arc or small pocket in angle-resolved photoemission (ARPES) on underdoped cuprates. 
As the observed intensity modulations of the spectrum are not compatible with the underlying tetragonal lattice symmetry we argue that this spectral-weight transfer appears to be accompanied by nematic order in the charge channel.

\section{Experiments And Results}
The single crystals of \naccod\ obtained by a flux method under high pressure \cite{Kohsaka_J.Am.Chem.Soc._2002_v124_p12275} were cut into thin ($d\sim 100\unit{nm}$) films using an ultramicrotome  under nitrogen atmosphere to avoid sample damage due to air-exposure. In the spectrometer the films were aligned \emph{in situ} with electron diffraction showing the high quality of our samples and allowing for polarization dependent investigations along well defined directions within the \cuopl. The measurements were carried out using a dedicated transmission electron energy-loss spectrometer \cite{Fink_Adv.Electron.ElectronPhys._1989} equipped with a helium flow cryostat employing a primary electron energy of $172\unit{keV}$ and energy and momentum resolutions of $\Delta E=80\unit{meV}$ and $\Delta q=0.035\rangstrom$, respectively.

\begin{figure*}
  \begin{center}
    \subfigure{%
      \includegraphics[scale=0.235]{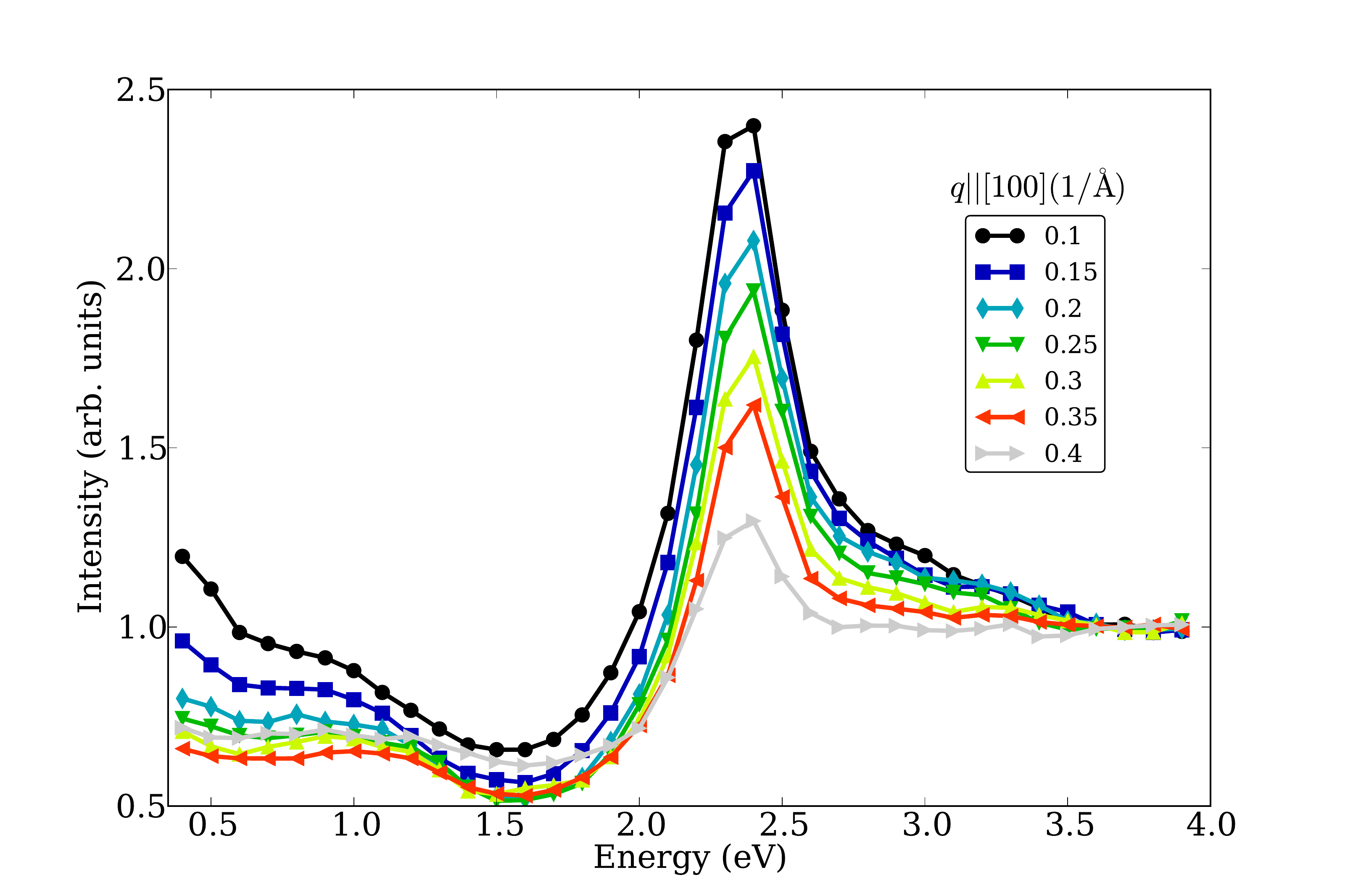}
      \label{fig:naccoc_x10_100_overview}}
    \subfigure{%
      \includegraphics[scale=0.235]{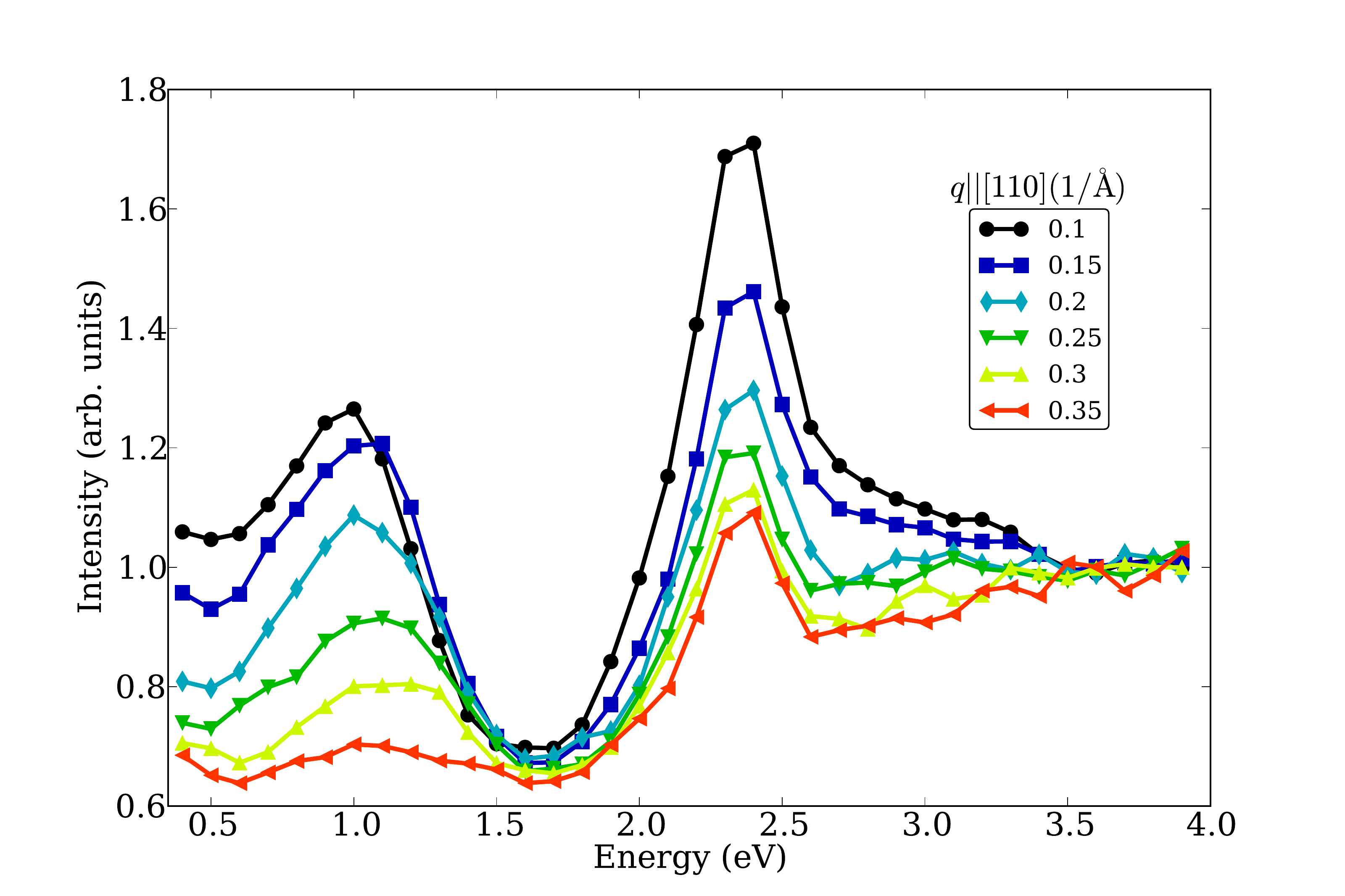}
      \label{fig:naccoc_x10_110_overview}}		
  \end{center}
  \caption{(Color Online) The EELS intensity in \naccod\ for momentum transfers parallel to $\direction{100}$ (left panel) and $\direction{110}$ (right panel) measured at room temperature. Note the intensity enhancement around $1\unit{eV}$ parallel to the $\direction{110}$ direction. All spectra have been normalized on the high-energy side between $3.5\unit{eV}$ and $4\unit{eV}$.}
  \label{fig:naccox_momentum_dependence}		
\end{figure*}

In \figref{fig:naccox_momentum_dependence} we present the behavior of the EELS intensity for \naccod\ in two high-symmetry directions within the \cuopl. The momentum evolution in the region of the CT peak around $2\unit{eV}$ were investigated in a previous report \cite{Schuster_Phys.Rev.B_2009_v79_p214517} and we therefore focus on the energy range around $1\unit{eV}$ in the following. As can be seen for momentum transfers parallel to the $\direction{110}$ direction there is a well-pronounced peak in the tail of the zero-loss line which disperses to higher energies upon leaving the center of the Brillouin zone (BZ). This peak is, however, strongly suppressed for $q\|\direction{100}$ (along the copper-oxygen bonds). Note that this anisotropy is absent in the insulating \scoc\  \cite{Neudert_PhysicaB:CondensedMatter_1997_v230-232_p847}, in the strongly underdoped \naccoc\ \cite{Schuster_Phys.Rev.B_2009_v79_p214517} and also the optimally doped \bscco\ \cite{Nucker_Phys.Rev.B_1991_v44_p7155}.

\begin{figure}
  \centering%
  \includegraphics[scale=.25]{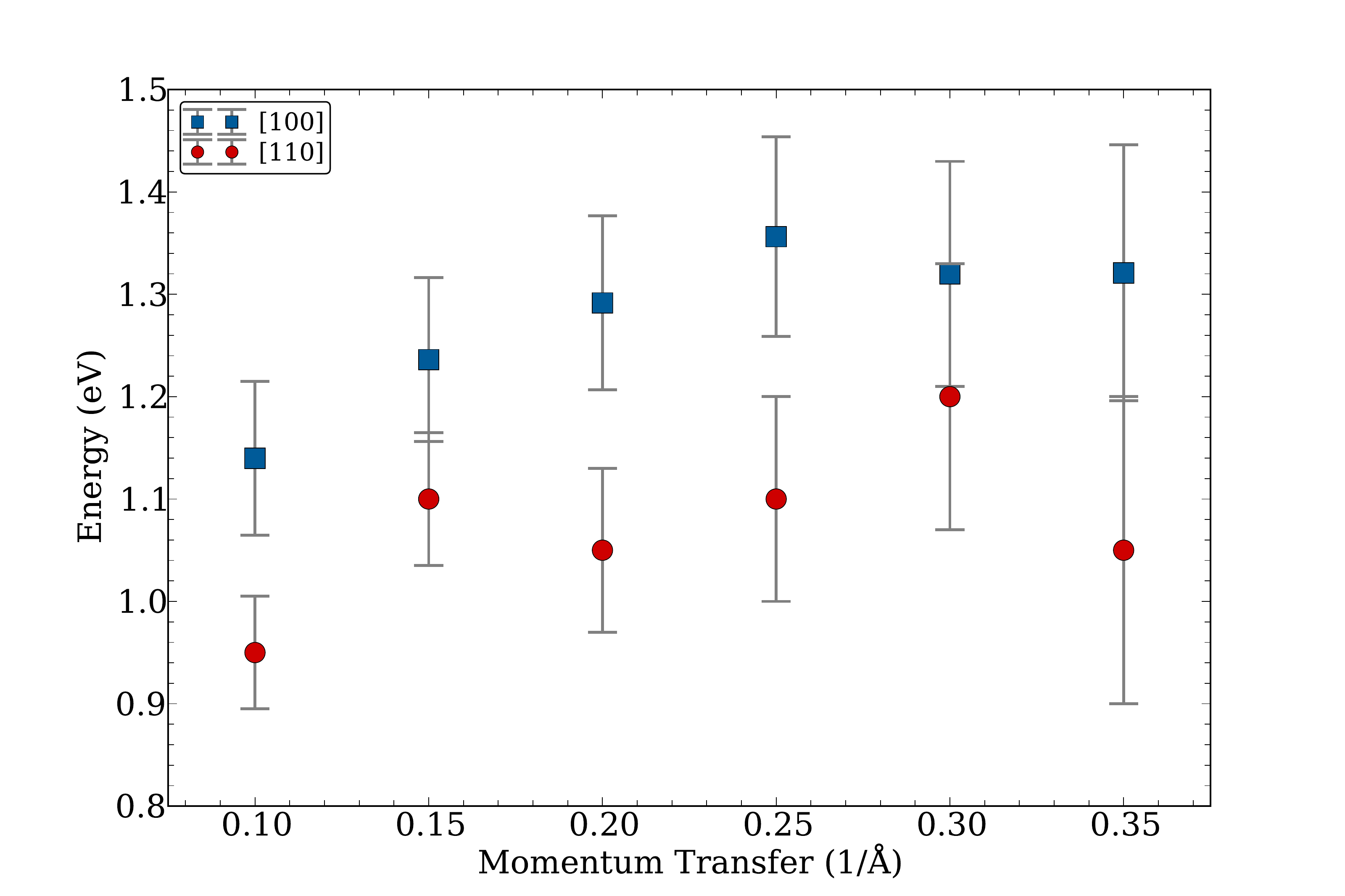}%
 \caption{(Color Online) The momentum dependence of the $1\unit{eV}$ feature seen in \figref{fig:naccox_momentum_dependence}. See text for details.}%
 \label{fig:naccoc_plasm_disp}%
\end{figure}

To further quantify the $1\unit{eV}$ feature we present its momentum evolution in \figref{fig:naccoc_plasm_disp} for both lattice directions. In order to obtain an unbiased estimate for the energetic position we did not remove the quasi-elastic line. While the energy values for $q\|\direction{110}$ shown in \figref{fig:naccoc_plasm_disp} track the position of the peak maximum around $1\unit{eV}$ this procedure could not be applied to the spectra in the \direction{100} direction as there the peak is hardly observable. Therefore we took the zero crossing of the second derivative between $1\unit{eV}$ and $1.5\unit{eV}$ as the characteristic feature for the spectra and present its momentum dependence in \figref{fig:naccoc_plasm_disp}. Consequently the different onset energies may be considered as an artifact of the data evaluation. As can be seen from this analysis, in both directions the dispersion is positive and has  a bandwidth of about $200\unit{meV}$.

\begin{figure*}
  \centering%
  \subfigure{%
    \includegraphics[scale=.45]{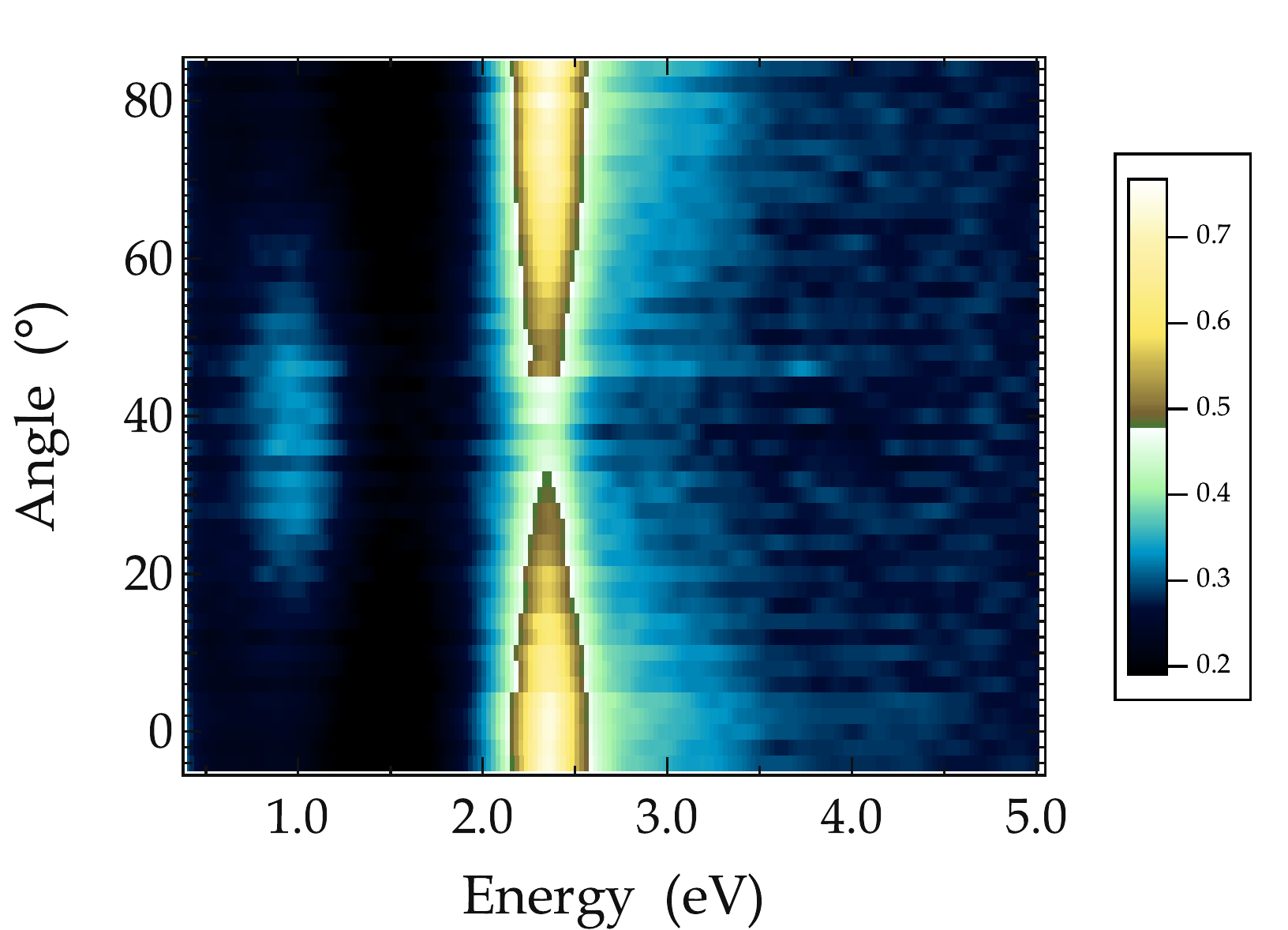}%
  }
  \subfigure{%
    \includegraphics[scale=.35]{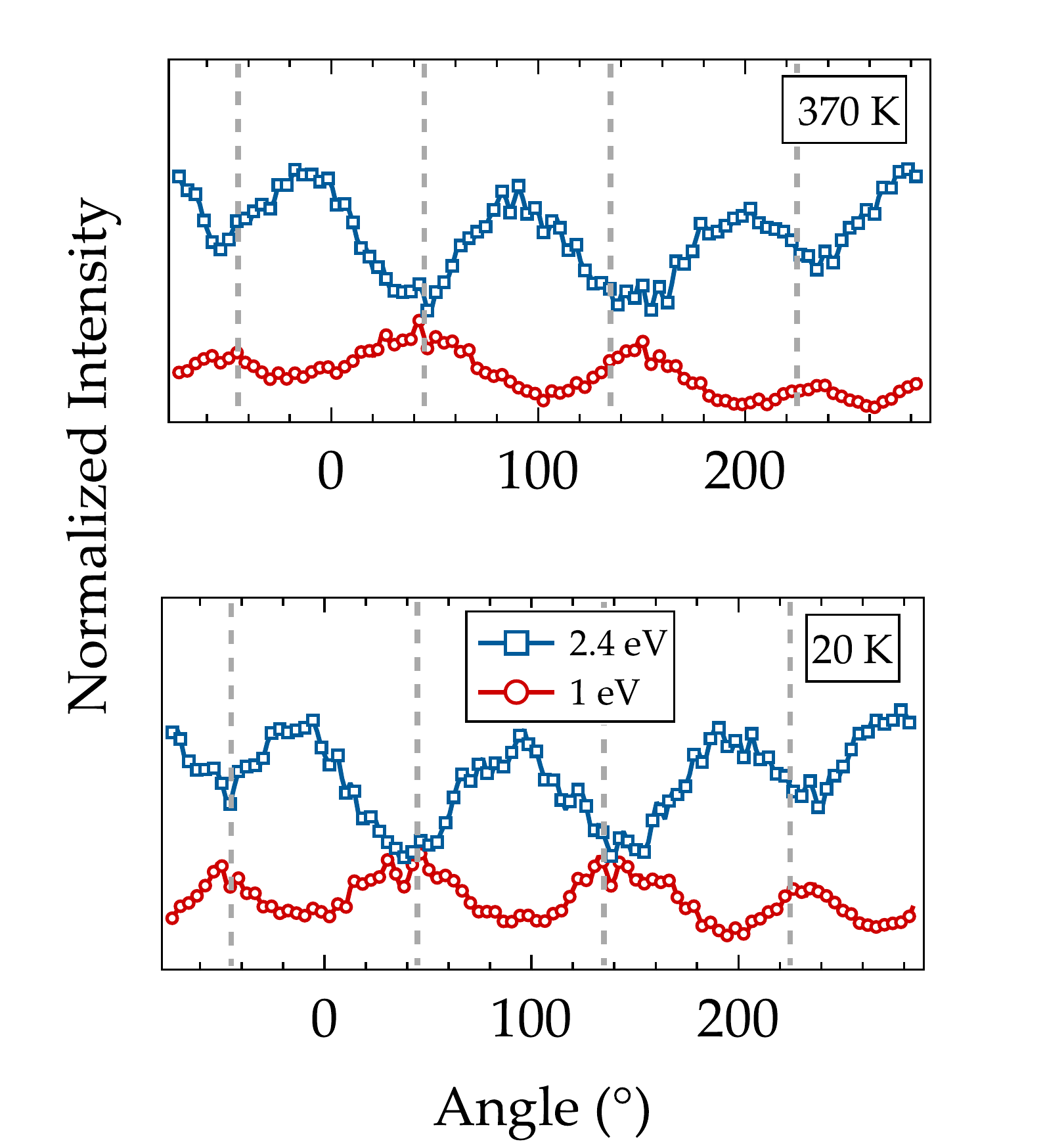}%
  }
  \caption{(Color Online) Left panel: Angular (polarization) dependence of the EELS intensity for $q=0.1\rangstrom$ measured at room temperature normalized on the high-energy side between $4\unit{eV}$ and $5\unit{eV}$.  Right panel: Constant energy cuts for $q=0.1\rangstrom$ around the full BZ (see text for details). Note the fourfold symmetry. The dashed bars correspond to the nodal directions as derived from the crystal structure measured with elastic scattering. In both panels the angle is measured relative to the $\direction{100}$ direction.}%
  \label{fig:exciton_plasmon_map_rt}%
\end{figure*}

To gain a deeper understanding we measured the EELS intensity for different temperatures and angles within the \cuopl. A summary of these results is presented in \figref{fig:exciton_plasmon_map_rt}. The left panel shows an angular map measured over one quadrant of the BZ at room temperature with a constant  momentum transfer of $q=0.1\rangstrom$. Obviously, there is a significant angular range where the $1\unit{eV}$ feature gains substantial weight at the expense of the CT peak. Though not perfectly symmetric \footnote{We attribute the angular mismatch to a slight spectrometer asymmetry.}, the portion of the BZ where the low-energy peak is most pronounced clearly corresponds to the so called nodal direction; the momentum region where the  superconducting and also the pseudogap (PG) are known to approach zero from ARPES \cite{Damascelli_Rev.Mod.Phys._2003_v75_p473}. 
The temperature (in)dependence of the discussed effect is summarized in the right panel of \figref{fig:exciton_plasmon_map_rt}. Those curves have been obtained by measuring angular cuts at constant energy-loss of $1\unit{eV}$ and $2.4\unit{eV}$. Subsequently, these two spectra were normalized by a similar one measured at $4\unit{eV}$ energy-loss. The angular independence of the EELS signal on the high-energy side allows us to consider this third spectrum as a suitable normalization background. Note, however, that the observed asymmetric signal does not depend on the energy range taken for the normalization. Clearly, the obtained pattern shows a fourfold symmetry over the entire angular range of the BZ with the maximum of the $1\unit{eV}$ peak always located around the nodal directions (indicated by the vertical dashed bars), in agreement with the left panel of \figref{fig:exciton_plasmon_map_rt}. From this we conclude that the observed effect is intrinsic and not caused by the sample preparation. Were this the case the preparation procedure would produce a distinguished axis parallel to the cutting direction of the sample, the result being a pattern with a periodicity of $\pi$ instead of the $\pi/2$-period we observe. In addition, the intensity redistribution between the features around $1\unit{eV}$ and $2\unit{eV}$ is surprisingly robust against temperature variations as we do not observe any changes of the periodicity or the amplitude between the highest and lowest measured temperatures shown in \figref{fig:exciton_plasmon_map_rt} (at least with the resolution accessible to us). The same holds true for all intermediate temperature steps investigated.

\begin{figure*}
  \centering%
  \includegraphics[scale=.425]{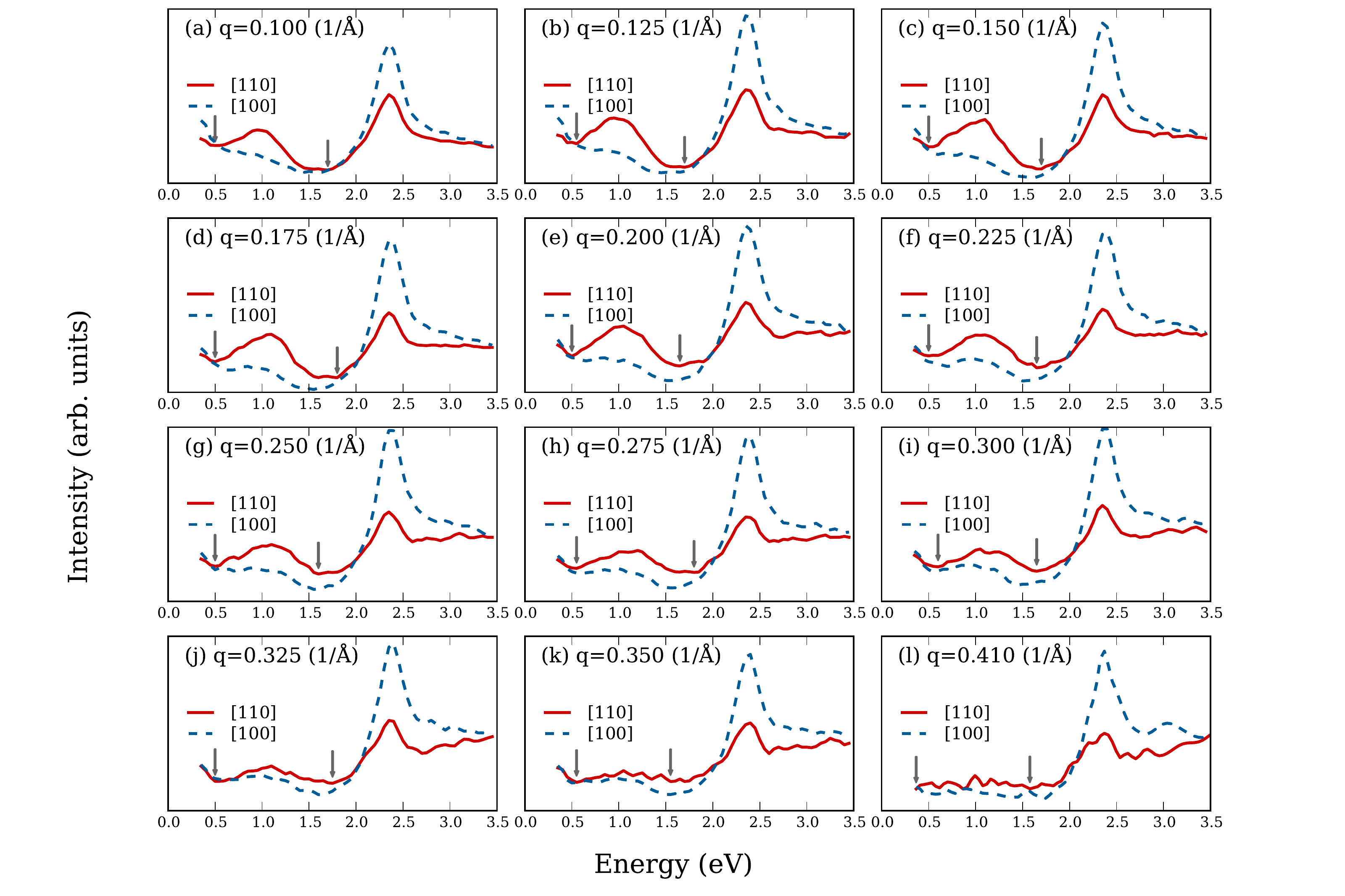}%
  \caption{(Color Online) The asymmetry of the EELS intensity for the two in-plane directions as a function of momentum. For each momentum value, the curves are normalized between $3.5\unit{eV}$ and $4\unit{eV}$ and the arrows indicate the integration ranges for the evaluation of  Eq.~\ref{eq:1}.}%
  \label{fig:intensity_variation_vs_q}%
\end{figure*}

We emphasize that we do not find evidence for superstructure reflections in the elastic scattering neither at high nor at low temperatures. This agrees with earlier x-ray scattering experiments on this system \cite{Smadici_Phys.Rev.B_2007_v75_p75104}. In addition the pattern seen in \figref{fig:exciton_plasmon_map_rt} changes quantitatively (see below) but not qualitatively  when changing the momentum transfer between $q=0.08\rangstrom$ and $q\sim0.4\rangstrom$ and we do not see any indication for a resonance-like enhancement of the effect at the momentum transfer of the checkerboard $q=2\pi/4a_0\approx0.41\rangstrom$ \cite{Hanaguri_Nature_2004_v430_p1001} (see \figref{fig:intensity_variation_vs_q}) with the lattice constant $a_0=3.85~\mbox{\AA}$.

In order to track the momentum evolution of the intensity anisotropy we plot the spectra along the \direction{100}- and \direction{110}-direction for a series of $q$-values in \figref{fig:intensity_variation_vs_q}. Already by visual inspection it is clear that the anisotropy persists but is not monotonic as a function of $q$. 

\begin{figure}
  \centering%
  \includegraphics[scale=.25]{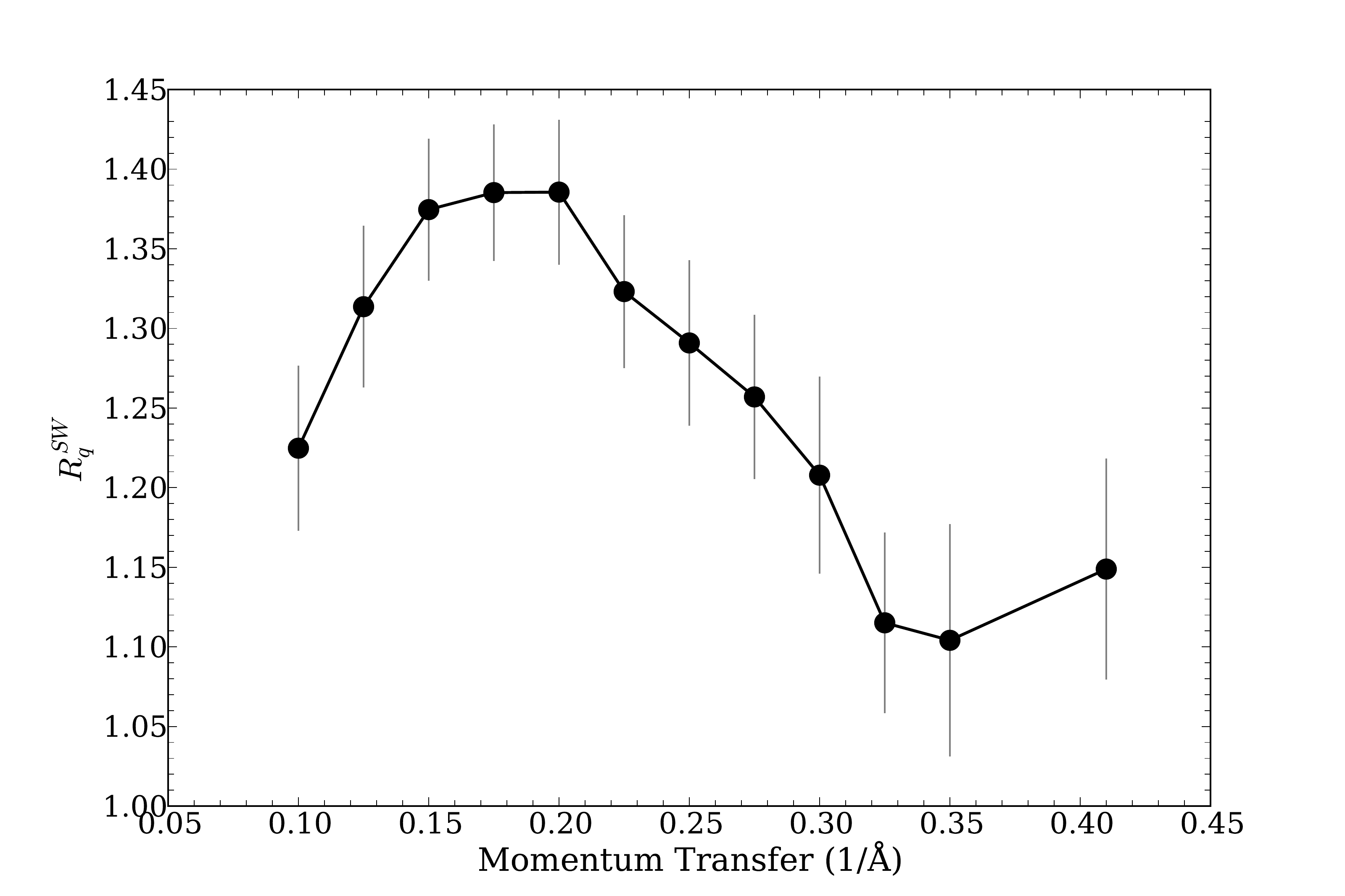}%
  \caption{
The momentum dependence of the ratio defined in Eq.~\ref{eq:1}.}%
  \label{fig:sw_asymmetry_vs_q}%
\end{figure}

As a more quantitative measure of the anisotropic EELS intensity we define the following ratio:
 \begin{align}
   R^{SW}_q&=\frac{SW_{q}^{\direction{110}}}{SW_{q}^{\direction{100}}}\label{eq:1}
\end{align}
where the 
``spectral weights'' $SW_q^i$ are determined by 
$SW_q^i=\int_{\omega_1}^{\omega_2}\mathrm{d}\omega\,\omega\,I (\omega,q)$ with the normalized intensities $I (\omega,q)$ shown in \figref{fig:intensity_variation_vs_q} and the momentum transfer $q$ parallel to the directions indicated by the superscript $i\in\left\{\direction{100},\direction{110}\right\}$. Note that if one converted the intensities $I(\omega,q)$ to the absolute value of \intextlf\ by means of a Kramers-Kronig (KK) transform the integrals $SW_q^i$ would indeed correspond to the spectral weight (and also be constrained by the $f$-sum rule \cite{Schnatterly1979}) which motivates our nomenclature. We did not, however, perform the KK-calculation, because, in particular for the \direction{100}-direction, the subtraction of the quasi-elastic line is highly ambiguous. The boundaries of the integration $\omega_1$ and $\omega_2$ correspond to the dips in the EELS intensity below and above the $1\unit{eV}$ feature along \direction{110}, respectively (see the vertical arrows in \figref{fig:intensity_variation_vs_q}). The evolution of the ratio defined in Eq.~\ref{eq:1}  as a function of momentum transfer is visualized in \figref{fig:sw_asymmetry_vs_q}. As can be seen, there is a rather well defined maximum around the incommensurate $q_c\sim0.175\rangstrom\ldots0.2\rangstrom$. 

\section{Discussion}
Although there is experimental evidence that the infrared optical response in underdoped cuprates contains more than a simple Drude term \cite{Waku_Phys.Rev.B_2004_v70_p134501,Lee_Phys.Rev.B_2005_v72_p54529} we attribute the intensity increase around $1\unit{eV}$ and in particular the well pronounced peak in the $\direction{110}$ direction to the Drude plasmon caused by the doped charge carriers. This is motivated by the metallic resistivity and the clearly visible plasma edge seen in the reflectivity of \naccod\ at this energy \cite{Waku_Phys.Rev.B_2004_v70_p134501}. In addition earlier EELS reports on  optimally doped cuprates, in particular from the Bi- and Y-families, found similar features that have been interpreted analogously \cite{Romberg_Z.Phys.B_1990_v78_p367,Wang_Phys.Rev.B_1990_v42_p420,Nucker_Phys.Rev.B_1991_v44_p7155,Nucker_Phys.Rev.B_1989_v39_p12379,Knupfer_PhysicaC_1994_v230_p121}. In contrast to our observation from \figref{fig:naccoc_plasm_disp}, however, all these experiments show a quadratic plasmon dispersion that follows the conventional behavior predicted for simple metals \cite{Raether_Excitationofplasmonsandinterbandtransitionsbyelectrons_1980} with a bandwidth that is larger than the one reported here by at least a factor of two. As the slope of the plasmon dispersion in a simple metal is proportional to the squared Fermi velocity $v_F\sim n^{\nicefrac{1}{3}}$ with the charge-carrier density $n$ the smaller dispersion in \naccod{} might simply reflect the (compared to the optimally-doped systems) smaller FS seen in ARPES \cite{Shen_Science_2005_v307_p901}. This is also in line with the systematic blue-shift of the plasma edge as a function of doping in \naccox{} \cite{Waku_Phys.Rev.B_2004_v70_p134501}.

We now turn our attention to the obvious anisotropy in the charge response for different directions within the \cuopl.

According to earlier reports \cite{Waku_Phys.Rev.B_2004_v70_p134501,Miller_Phys.Rev.B_1990_v41_p1921,Kohsaka_J.Am.Chem.Soc._2002_v124_p12275,Argyriou_Phys.Rev.B_1995_v51_p8434} but also from our elastic scattering data (not shown) the crystal structure of the \naccox\ system is perfectly tetragonal without any signs of buckling or orthorhombicity often found in other cuprate families. This implies that for $q=0$ the dielectric function---which is in the general case of anisotropic systems a tensor---is diagonal and contains only \emph{one} independent in-plane component and consequently should be \emph{iso}tropic throughout the entire \cuopl\ \cite{Landau_Electrodynamicsofcontinuousmedia_1984}. Therefore also $1/\epsilon(\omega,\bm{q})$, which we probe in EELS, should be describable by a single isotropic component. Although we cannot measure at $q\equiv0$ due to the increasing influence of the zero-loss peak and the enhancement of surface scattering \cite{Raether_Excitationofplasmonsandinterbandtransitionsbyelectrons_1980}, in general the momentum transfer of $q=0.1\rangstrom$, which corresponds to only about 6\% of the BZ size, is small enough to be considered as the optical limit ($q=0$) (see also the isotropy for this values of momentum transfer in the response of other cuprates \cite{Neudert_PhysicaB:CondensedMatter_1997_v230-232_p847,Schuster_Phys.Rev.B_2009_v79_p214517,Nucker_Phys.Rev.B_1991_v44_p7155}). Consequently, we attribute the behavior described above to an intrinsic (electronic) breaking of the underlying tetragonal lattice symmetry as our data imply strong polarization-dependent changes of the charge response, equivalent to two in-plane components of the dielectric tensor even in the limit $q\rightarrow 0$. This is compatible with the STM data \cite{Kohsaka_Science_2007_v315_p1380,Kohsaka_NatPhys_2012_v8_p534} where evidence is found for a local breaking of the symmetry from $\mathcal{C}_4$ (tetragonal) to $\mathcal{C}_2$ (orthorhombic) which can be ascribed to the presence of a nematic order parameter \cite{Vojta_Adv.Phys._2009_v58_p699,Fradkin_Annu.Rev.Condens.MatterPhys._2010_v1_p153}. Note, however, that while the STM data implies a higher metallic character parallel to the bonds our data seem to indicate an enhanced density of states along the diagonals of the unit cell (BZ). %
Our data therefore also appears at odds with results based on Raman- \cite{Tassini_Phys.Rev.B_2008_v78_p20511,Tassini_Phys.Rev.Lett._2005_v95_p117002} and neutron-scattering \cite{Wakimoto_Phys.Rev.B_1999_v60_p769} where stripes are along the diagonal only for a doping too low to induce superconductivity. There are several possible scenarios to reconcile the observed $\pi/2$-periodic signal (see \figref{fig:exciton_plasmon_map_rt}). Either there are perpendicular 1D domains within a single \cuopl\ or the charge modulation is 2D. Another possibility is that different \cuopl s along the path of the electron beam have perpendicular 1D domains and the observed intensity pattern is just a superposition of the contribution of all unit-cells in the sample. In any case, along the domains there is metallic transport reflected by the appearance of the plasmon whereas perpendicular to them the spectrum is dominated by the CT excitations reminiscent of the insulator. These domains are highly fluctuating but the timescale of the electron-scattering process is short enough that EELS measures a snapshot of this dynamic behavior. At low enough temperatures these charge fluctuations may lock in to form a well-ordered pattern but due to the transmission geometry of the experiment we are limited to $T\geqslant20\unit{K}$. From the absence of superstructure reflections and the robustness of the tetragonal lattice symmetry, we conjecture that \naccod\ tends to show nematic order. 
Our data therefore also indicate that the regular checkerboard pattern seen in the STM \cite{Hanaguri_Nature_2004_v430_p1001} may be pinned by the surface, in agreement with  the findings of \intextcite{Brown_Phys.Rev.B_2005_v71_p224512}. From \figref{fig:sw_asymmetry_vs_q} we identify a characteristic  momentum $q_c\sim0.175\rangstrom\ldots0.2\rangstrom$ of the nematic fluctuations which translates to a length scale of $l_c=\frac{2\pi}{q_c}\sim30\,\text{\AA}\ldots35\,\text{\AA}\sim8a_0\ldots9 a_0$.


From the ARPES literature on underdoped \naccox\ \cite{Shen_Science_2005_v307_p901,Ronning_Phys.Rev.B_2005_v71_p94518} and also other cuprate systems \cite{Damascelli_Rev.Mod.Phys._2003_v75_p473} it is known that the large FS observed in the optimally and overdoped compounds is confined to a small angular range---either in the form of an arc or closed pocket \cite{Meng_Nature_2009_v462_p335}---around the nodal direction which increases with doping.  From our data, in particular from \figref{fig:exciton_plasmon_map_rt} it is clear where these metallic states derive from: they are created by a transfer of spectral weight from the upper Hubbard band (the excitations around $2.5\unit{eV}$ in \figref{fig:exciton_plasmon_map_rt}) down to low energies in the vicinity of the Fermi level. In EELS these low-lying states show up at finite energies (around $1\unit{eV}$) because for a metal a peak in $\Im(-1/\epsilon)=\epsilon_2/(\epsilon_1^2+\epsilon_2^2)$ occurs at the plasma frequency which is determined by the density of free charges and therefore finite. Note that such a spectral-weight transfer is considered to be a hallmark of strongly-correlated systems \cite{Meinders_Phys.Rev.B_1993_v48_p3916}. Angle-resolved \emph{inverse} photoemission could reveal the corresponding behavior in the single-particle channel.

Interestingly, a similar anisotropic appearance of metallic states upon doping is also predicted theoretically for the Hubbard model by cluster extensions of dynamical mean-field theory \cite{Civelli_Phys.Rev.Lett._2005_v95_p106402} and variational cluster approaches \cite{Arrigoni_NewJ.Phys._2009_v11_p55066}.

As the spectral-weight transfer, however, is expected to occur
symmetrically for all four nodal points of the BZ and therefore
preserves the underlying tetragonal symmetry of the lattice our observations seem to require the collaborative action of (fluctuating) charge order \emph{and} spectral-weight transfer.

\section{Summary}
To summarize we investigated the energy and momentum dependent charge response in \naccod{} by means of electron energy-loss spectroscopy. We find a polarization dependent transfer of spectral weight between the CT excitations---dominating the spectrum around $(\pi,0)$---and the metallic plasmon (around $(\pi,\pi)$) which can be understood as an anisotropic metal-insulator transition within the BZ. The observed pattern of the charge response indicates the breaking of the underlying tetragonal lattice symmetry which we take as indicative for a nematic order parameter in the \cuopl. In addition, the plasmon is found to disperse with a bandwidth smaller than in optimally doped cuprates. All effects are robust against temperature variation arguing in favor of a highly fluctuating order.

\section{Acknowledgments}
We appreciate experimental support by R. Hübel, R. Schönfelder and S. Leger and stimulating discussions with J.\,v. Wezel and J.\,v.\,d. Brink. This project is supported by the DFG through KN393/13 and by the Grants-in-Aid for Scientific Research, MEXT of Japan under Grants No.
17105002, No. 19014010, and No. 19340098.

%

\end{document}